\documentclass{article}
\usepackage[utf8]{inputenc}
\usepackage{amsmath,graphicx}
\usepackage{INTERSPEECH2019}
\usepackage{multirow}
\usepackage{adjustbox}
\usepackage{dblfloatfix}
\usepackage{url,times,booktabs}
\usepackage{comment}
\usepackage{tabularx}
\usepackage{makecell}
\usepackage{cite}
\usepackage{enumerate}
\usepackage{amsmath,graphicx,url,times,booktabs, tabularx}
\usepackage{arydshln}

\title{RawNet: Advanced end-to-end deep neural network using raw waveforms\\for text-independent speaker verification}
\name{Jee-weon Jung, Hee-Soo Heo, Ju-ho Kim, Hye-jin Shim, and Ha-Jin Yu$^\dag$\thanks{$^\dag$ Corresponding author}\thanks{This work was supported by the Technology Innovation Program (10076583, Development of free-running speech recognition technologies for embedded robot system) funded By the Ministry of Trade, Industry \& Energy(MOTIE, Korea)}}
\address{
  School of Computer Science, University of Seoul, South Korea}
\email{jeewon.leo.jung@gmail.com,
  zhasgone@naver.com,
  wngh1187@naver.com,
  shimhz6.6@gmail.com,
  hjyu@uos.ac.kr}

\begin{document}
\maketitle
\begin{abstract}
Recently, direct modeling of raw waveforms using deep neural networks has been widely studied for a number of tasks in audio domains. 
In speaker verification, however, utilization of raw waveforms is in its preliminary phase, requiring further investigation. 
In this study, we explore end-to-end deep neural networks that input raw waveforms to improve various aspects: front-end speaker embedding extraction including model architecture, pre-training scheme, additional objective functions, and back-end classification. 
Adjustment of model architecture using a pre-training scheme can extract speaker embeddings, giving a significant improvement in performance. 
Additional objective functions simplify the process of extracting speaker embeddings by merging conventional two-phase processes: extracting utterance-level features such as i-vectors or x-vectors and the feature enhancement phase, e.g., linear discriminant analysis. 
Effective back-end classification models that suit the proposed speaker embedding are also explored. 
We propose an end-to-end system that comprises two deep neural networks, one front-end for utterance-level speaker embedding extraction and the other for back-end classification. 
Experiments conducted on the \texttt{VoxCeleb1} dataset demonstrate that the proposed model achieves state-of-the-art performance among systems without data augmentation. 
The proposed system is also comparable to the state-of-the-art x-vector system that adopts data augmentation.
\end{abstract}


\section{Introduction}
Direct modeling of raw waveforms using deep neural networks (DNNs) is increasingly prevalent in a number of tasks due to advances in deep learning \cite{SpeechRACNN, sainath2015learning, SpeechRACNN2, spoofCLDNN, jung2018complete, muckenhirn2018towards, jung2018avoiding22, RACNN}. 
In speech recognition, studies such as those of Palaz \textit{et al.}, Sainath \textit{et al.}, and Hoshen \textit{et al.} deal with raw waveforms as input \cite{SpeechRACNN, sainath2015learning, SpeechRACNN2}. 
In speaker recognition, studies by Jung \textit{et al.} and Muckenhirn \textit{et al.} were the first to comprise systems that input raw waveforms \cite{jung2018complete, muckenhirn2018towards, jung2018avoiding22}. 
Other domains such as spoofing detection and automatic music tagging are also adopting raw waveform inputs \cite{spoofCLDNN, RACNN}. 

DNNs that directly input raw waveforms have a number of advantages over conventional acoustic feature-based DNNs. 
First, minimization of pre-processing removes the need for exploration of various hyper-parameters such as the type of acoustic feature to use, window size, shift length, and feature dimension. 
This is expected to lower entry barriers to conducting studies and lessen the burden of follow-up studies. 
%Moreover, raw waveform DNNs are generally applicable to a number of audio domain tasks with minimal alteration. 
Additionally, with recent trends of DNN replacing more sub-processes in various tasks, a raw waveform DNN is well positioned to benefit from future advances in deep learning. 

Studies across various tasks have shown that an assembly of multiple frequency responses can be extracted when raw waveforms are processed by each kernel of convolutional layers \cite{muckenhirn2018towards, ravanelli2018speaker}. 
Spectrograms, compared to raw waveform DNNs, have linearly positioned frequency bands, meaning the first convolutional layer sees only adjacent frequency bands (although repetition of convolutions can aggregate various frequency responses at deeper layers). 
In other words, spectrogram-based CNN can see fixed frequency regions depending on the internal pooling rule. 
This difference is hypothesized to increase the potential of the directly modeling raw waveforms; as increasing amounts of data become available, this data-driven approach can extract an aggregation of informative frequency responses appropriate to the target task. 
%Unlike spectrograms with linearly positioned frequency bands, where the convolutional layer at first sees only adjacent frequency bands, raw waveforms can present an aggregation of various frequency responses in a data-driven manner dependent on the target task. 

In this study, we improve various aspects of the raw waveform DNN proposed by Jung \textit{et al.}, which was the first end-to-end model in speaker verification using raw waveforms \cite{jung2018complete, jung2018avoiding22}. 
This model extracts frame-level embeddings using residual blocks with convolutional neural networks (CNN) \cite{Residual, full_pre}, and then aggregates features into utterance level using long short-term memory (LSTM) \cite{LSTM, LSTM2}. 
Key improvements made by our study include the following:
\begin{enumerate}
    \item Model architecture: adjustments to various network configurations
    \item CNN pre-training scheme: removal of inefficient aspects of the multi-step training scheme in \cite{heo2017joint, jung2018avoiding22}
    \item Objective function: additional objective functions to incorporate speaker embedding extraction and feature enhancement phase
    \item Back-end classification models: comparison of various DNN-based back-end classifiers and proposal of a simple, effective back-end DNN classifier
\end{enumerate}
Through changing these aspects, performance is significantly enhanced. 
The equal error rate (EER) of utterance-level speaker embedding DNN with cosine similarity on the \texttt{VoxCeleb1} dataset is 4.8 \%, showing 44.8 \% relative error rate reduction (RER) compared to the baseline \cite{jung2018avoiding22}. 
An EER of 4.0 \% was achieved for the end-to-end model using two DNNs, showing an RER of 46.0 \%.

The rest of this paper is organized as follows. 
Section 2 describes the front-end speaker embedding extraction model. 
Section 3 addresses various back-end classification models. 
Experiments and results are in Sections 4 and 5 and the paper is concluded in Section 6.

\section{Front-end: RawNet}
We propose a model (referred to as ``RawNet'' for convenience) that is an improvement of the CNN-LSTM model in \cite{jung2018complete, jung2018avoiding22} by changing architectural details (Section 2.1.), proposing a modified pre-training scheme (Section 2.2.), and incorporating a speaker embedding enhancement phase (Section 2.3.). 

\subsection{Model architecture}
The DNN used in this study comprises residual blocks, a gated recurrent unit (GRU) layer \cite{cho2014learning, chung2014empirical}, a fully-connected layer (used for extraction of speaker embedding), and an output layer. 
In this architecture, input features are first processed using the residual blocks \cite{Residual} to extract frame-level embeddings. 
The residual blocks comprise convolutional layers with identity mapping \cite{full_pre} to facilitate the training of deep architectures. 
A GRU is then employed to aggregate the frame-level features into a single utterance-level embedding. 
Utterance-level embedding is then fed into one fully-connected layer. 
The output of the fully-connected layer is used as the speaker embedding and is connected to the output layer, where the number of nodes is identical to the number of speakers in the training set. 
The proposed RawNet architecture is depicted in Table 1.

It includes a number of modifications to the CNN-LSTM model in \cite{jung2018complete, jung2018avoiding22}, allowing for further improvement. 
First, activation functions are changed from rectified linear units (ReLU) to leaky ReLU. 
Second, the LSTM layer is changed to a GRU layer. 
Third, the number of parameters is significantly decreased, including lower dimensionality of speaker embedding (from 1024 to 128).

\subsection{CNN pre-train scheme}
Extracting utterance-level speaker embeddings directly from raw waveforms often leads to overfitting toward the training set \cite{jung2018avoiding22}. 
In \cite{jung2018avoiding22}, multi-step training proposed in \cite{heo2017joint} was used to avoid such phenomenon. 
% This training scheme first trains a CNN (for frame-level training), and then expands to a CNN-LSTM (for utterance level) instead of training a CNN-LSTM with random initialization.
This training scheme first trains a CNN (for frame-level training), and then expands to a CNN-LSTM (for utterance level).
This scheme demonstrates significant improvement compared to training a CNN-LSTM with random initialization. 

% However, the multi-step training approach in \cite{jung2018avoiding22} contains inefficiency because a number of residual convolutional blocks and fully-connected layers are removed when expanding the trained CNN model to a CNN-LSTM model. 
However, the multi-step training approach in \cite{jung2018avoiding22} is inefficient because after training 9 residual blocks, 3 residual blocks which contains a number of layers are removed when expanding the trained CNN model to a CNN-LSTM model. 
In our study, a new approach of interpreting the CNN training phase as pre-training is applied. 
This approach adopts fewer residual convolutional blocks, i.e. 6, connected to a global average pooling layer. 
After training the CNN, only the global average pooling layer is removed. 
% In contrast, conventional approach in \cite{jung2018avoiding22} adopts a larger number of convolutional blocks, i.e. 9, then removes latter three blocks after training the CNN. 
The objective is to consider the number of convolutional blocks appropriate for training with the recurrent layer and not remove any parameters. 
This modification enables more efficient and faster training. 
Application of model architecture modifications detailed in Section 2.1, and the CNN pre-training scheme exhibited an RER of 26.4 \% (see Table 2).

\begin{table}[t]
 \caption{RawNet architecture. For convolutional layers, numbers inside parentheses refer to filter length, stride size, and number of filters. For gated recurrent unit (GRU) and fully-connected layers, numbers inside the parentheses indicate the number of nodes. An input sequence of 59,049 is based on the training mini-batch configuration. At the evaluation phase, input sequence length differs. Center loss and between-speaker loss is omitted for simplicity. For residual blocks, layers under the dotted line are conducted after residual connection.}
  \centering
  \label{tab:table1}
  \begin{tabular}{l c c}
  \Xhline{2\arrayrulewidth}
  \textbf{Layer} & \textbf{Input:59,049 samples} & \textbf{Output shape}\\
  \Xhline{2\arrayrulewidth}
  \multirow{2}{*}{Strided} & Conv(3,3,128) & \multirow{3}{*}{(19683, 128)}\\
  \multirow{2}{*}{-conv}& BN & \\
  & LeakyReLU & \\
  \midrule
  Res block & 
    %\[ \left \{
    $\left \{
      \begin{tabular}{c}
      Conv(3,1,128)\\
      BN \\
      LeakyReLU\\
      Conv(3,1,128)\\
      BN\\
      \hdashline
      LeakyReLU\\
      MaxPool(3)\\
      \end{tabular}
    \right \}$
    %\]
    $\times$2
    
  & (2187, 128)\\
  \midrule
  Res block & 
    
    $\left \{
      \begin{tabular}{c}
      Conv(3,1,256)\\
      BN \\
      LeakyReLU\\
      Conv(3,1,256)\\
      BN\\
      \hdashline
      LeakyReLU\\
      MaxPool(3)\\
      \end{tabular}
    \right \}$ 
    $\times$4
  & (27, 256)\\
  \midrule
  GRU & GRU(1024) & (1024,)\\
  \midrule
  Speaker & \multirow{2}{*}{FC(128)} & \multirow{2}{*}{(128,)}\\
  embedding & & \\
  \midrule
  Output & FC(1211) & (1211,)\\
  \bottomrule
  \end{tabular}
\end{table}

\subsection{Additional objective functions for speaker embedding enhancement}
For speaker verification, a number of studies enhance extracted utterance-level features through an additional process before back-end classification. 
Linear discriminant analysis (LDA) in i-vector/PLDA systems is one example \cite{dehak2011front, PLDA}. 
Well-known methods such as LDA or recent DNN-based deep embedding enhancement, including discriminative auto-encoder (DCAE), have been applied for this purpose in feature enhancement. 
In such approaches, one of the main objectives is to minimize intra-class covariance and maximize inter-class covariance of utterance-level features. 
%In this study, we aim to directly extract an utterance-level speaker embedding that contains properties of conventional feature enhancement techniques, without additional sub-processes. 
In this study, we aim to incorporate two phases of speaker embedding extraction and feature enhancement into a single phase, using two additional objective functions. 

To consider both inter-class and intra-class covariance, we utilize center loss \cite{wen2016discriminative} and speaker basis loss \cite{heo2019end} in addition to categorical cross-entropy loss for DNN training. 
We adopt center loss \cite{wen2016discriminative} to minimize intra-class covariance while the embedding in the last hidden layer remains discriminative. 
To achieve this goal, center loss function was proposed as 
\begin{equation}
    \mathcal{L}_C = \frac{1}{2}\sum_{i=1}^{N}||x_i - c_{y_{i}}||^2_2,
\end{equation}
where $x_i$ refers to embedding of the $ith$ utterance, $c_{y_i}$ refers to the center of class $y_i$, and $N$ refers to the size of a mini-batch. 

Speaker basis loss \cite{heo2019end}, aims to further maximize inter-class covariance. 
This loss function considers a weight vector between the last hidden layer and a node of the softmax output layer as a basis vector for the corresponding speaker and is formulated as:

\begin{equation}
\mathcal{L}_{BS}=\sum\limits_{i=1}^{M}\sum\limits_{j=1,j \neq i}^{M}cos(w_{i},w_{j}),
\end{equation}
where $w_i$ is the basis vector of speaker $i$ and $M$ is the number of speakers within the training set. 
Hence, the final objective function used in this study is 

\begin{equation}
    \mathcal{L} = \mathcal{L}_{CE} + \lambda\mathcal{L}_C +
    \mathcal{L}_{BS},
\end{equation}
where $\mathcal{L}_{CE}$ refers to categorical cross-entropy loss and $\lambda$ refers to the weight of $\mathcal{L}_C$.

\begin{table}[t]
 \caption{Experimental results showing the effectiveness of the RawNet system. \textit{Intermediate} refers to results of application of the modifications proposed in Sections 2.1. and 2.2. The back-end classifier is fixed to cosine similarity.}
  \centering
  \label{tab:table1}
  \begin{tabular}{l c c c}
  \Xhline{2\arrayrulewidth}
  System & \textbf{Loss} & \textbf{Enhance} & \textbf{EER \%}\\
  \Xhline{2\arrayrulewidth}
  Baseline \cite{jung2018avoiding22} & Soft & None & 8.7\\
  \hline
    \multirow{2}{*}{\textit{Inter-}} & \multirow{3}{*}{Soft}& None & 6.8\\
   \multirow{2}{*}{\textit{mediate}}& & LDA & 6.5\\
  & & \textbf{DCAE} & \textbf{6.4}\\

  \midrule
  \multirow{3}{*}{RawNet}& \multirow{3}{*}{Soft+Center+BS}& \textbf{None} & \textbf{4.8}\\
   & & LDA & 7.6\\
   & & DCAE & 4.9\\

  \bottomrule
  \end{tabular}
\end{table}

\begin{table}[t]
 \caption{Comparison of equal error rate (EER) of various front-end speaker embedding extraction systems. The back-end classifier is fixed to cosine similarity.}
  \centering
  \label{tab:table1}
  \begin{tabular}{l c}
  \Xhline{2\arrayrulewidth}
  \textbf{System} & \textbf{EER \%}\\
  \Xhline{2\arrayrulewidth}
  i-vector & 13.8\\
  i-vector/LDA & 7.25\\
  \hline
  x-vector(w/o augment) \cite{shon2018frame}& 11.3\\
  x-vector(w augment) \cite{shon2018frame} & 9.9\\
  \hline
  \textbf{RawNet} & \textbf{4.8}\\
  \bottomrule
  \end{tabular}
\end{table}

\section{DNN-based back-end classification}
In speaker verification, cosine similarity and PLDA are widely used for back-end classification to determine whether two speaker embeddings belong to the same speaker \cite{PLDA}. 
Although PLDA has shown competitive results in a number of studies, DNN-based classifiers have also shown potential in previous researches \cite{jung2018complete}. 
A number of DNN-based back-end classifiers have been explored: concatenation of speaker embeddings, and b-vector and rb-vector systems \cite{b-vector, heo2016advanced}. 
A novel back-end classifier using DNN is introduced based on analysis of a b-vector system. 

The b-vector proposed by Lee \textit{et al.} exploits element-wise binary operation of speaker embeddings to represent relationships \cite{b-vector}. 
Operations include addition, subtraction, and multiplication.
The results are concatenated, composing a b-vector that has three times the dimensions of a single speaker embedding. 
Although the technique itself is simple, results demonstrate that binary operations effectively represent the relationship between the speaker embeddings. 

The rb-vector is an expansion of the b-vector, where an additional r-vector is used with the b-vector \cite{heo2016advanced}. 
The main purpose of the r-vector approach is to represent the relationship of the speaker embeddings to the training set. 
To compose r-vectors, a fixed number of representative vectors are derived using k-means on the speaker embeddings of the training set. 
For every trial, b-vectors are composed by conducting binary operations between representative vectors and the speaker embedding. 
These b-vectors are dimensionally reduced using principal component analysis (PCA) and then concatenated, producing r-vectors. 

The b-vector approach uses various element-wise binary operations to derive the relationship between the speaker embeddings. 
However, because weighted summation operations in a DNN can replace (or even find better combinations of) addition and subtraction, we hypothesized that the core term contributing to the success of the b-vector is the multiplication operation. 
Therefore, we propose an approach using the concatenation of the speaker embedding, test utterance, and their element-wise multiplication. 
Experimental results show that by only adding element-wise multiplication, performance exceeds that of the b-vector (see Table 4, `concat\&mul’).

\begin{table}[t]
 \caption{Comparison of various back-end classifiers. Speaker embeddings extracted from RawNet are used. Results are reported in terms of equal error rate (EER, \%)}  
  \centering
  \label{tab:table1}
  \begin{tabular}{l c }
  \Xhline{2\arrayrulewidth}
  \textbf{Classifier} & \textbf{EER \%}\\
  \Xhline{2\arrayrulewidth}
  Cosine similarity & 4.8\\
  \midrule
  PLDA & 4.8\\
  %PCA/PLDA & 4.7\\
  LDA/PLDA & 4.8\\
  \midrule
  b-vector & 4.1\\
  rb-vector & 4.1\\
  \textbf{concat\&mul} & \textbf{4.0}\\
  \bottomrule
  \end{tabular}
\end{table}

\begin{table*}[t]
 \caption{Overall results of speaker verification system application to the VoxCeleb 1 dataset.}  
  \centering
  \label{tab:table1}
  \begin{tabular}{l c c c c c c c }
  \Xhline{2\arrayrulewidth}
    & Input Feature & Front-end & Back-end & Loss & Dims & Augment & EER (\%) \\
  \Xhline{2\arrayrulewidth}
  Shon \textit{et al.} \cite{shon2018frame} & MFCC & x-vector & PLDA & Softmax & 600 &  Light & 6.0\\
  Shon \textit{et al.} \cite{shon2018frame} & MFCC & 1D-CNN & PLDA & Softmax & 512 &  Light & 5.3\\
  Hajibabaei \textit{et al.} \cite{hajibabaei2018unified} & Spectrogram & ResNet-20 & N/A & A-Softmax & 128 & Heavy & 4.4 \\
  Hajibabaei \textit{et al.} \cite{hajibabaei2018unified} & Spectrogram & ResNet-20 & N/A & AM-Softmax & 128 & Heavy & 4.3 \\
  Okabe \textit{et al.} \cite{okabe2018attentive} & MFCC & x-vector & PLDA & Softmax & 1500 & Heavy & \textbf{3.8}\\
  \Xhline{2\arrayrulewidth}
  
  Nagrani \textit{et al.} \cite{Voxceleb} & MFCC & i-vector & PLDA & - & - & - & 8.8\\
  \textbf{Ours} & MFCC & i-vector & PLDA & - & 250 & - & 5.1\\
  \hline
  Nagrani \textit{et al.} \cite{Voxceleb} & Spectrogram & VGG-M & Cosine & Metric learning & 256 & - & 7.8\\
  Jung \textit{et al.} \cite{jung2018avoiding22} & Raw waveform & CNN-LSTM & Cosine & Softmax & 1024 & - & 8.7\\
  Jung \textit{et al.} \cite{jung2018avoiding22} & Raw waveform & CNN-LSTM & b-vector & Softmax & 1024 & - & 7.7\\
  Shon \textit{et al.} \cite{shon2018frame} & MFCC & x-vector & PLDA & Softmax & 512 &  - & 7.1\\
  Shon \textit{et al.} \cite{shon2018frame} & MFCC & 1D-CNN & PLDA & Softmax & 600 &  - & 5.9\\

  \hline
  \textbf{Ours} & Raw waveform & RawNet & cosine & Softmax+Center+BS & 128 & - & 4.8\\
  \textbf{Ours} & Raw waveform & RawNet & concat\&mul & Softmax+Center+BS & 128 & - & \textbf{4.0}\\
  \Xhline{2\arrayrulewidth}
  \end{tabular}
\end{table*}

\section{Experimental settings}
Experiments in this study were conducted using Keras, a deep learning library in Python with Tensorflow back-end \cite{keras, tensorflow, tensorflow2}. 
Code used for experiments is available at \texttt{https://github.com/Jungjee/RawNet}.

\subsection{Dataset}
We use \texttt{VoxCeleb1} dataset which comprises approximately 330 hours of recordings from 1251 speakers in text-independent scenarios and has a number of comparable recent studies in the literature. 
All utterances are encoded at a 16 kHz sampling rate with 16-bit resolution. 
As the dataset comprises various utterances of celebrities from YouTube, it includes diverse background noise and varied duration.  
We followed the official guidelines which divide the dataset into training and evaluation sets of 1211 and 40 speakers respectively. 

\subsection{Experimental configurations}
We didn't apply any pre-processing, such as normalization, except pre-emphasis \cite{Pre-emphasis} to raw waveforms which were experimentally shown effective in our internal comparison experiments. 
For mini-batch construction, utterances were either cropped or duplicated (concatenated until the length reach) into 59049 samples ($\approx 3.59 s$) in the training phase, following \cite{jung2018complete, jung2018avoiding22}. 
In the evaluation phase, no adjustments were made to length; the whole utterance was used. 

RawNet comprises one strided convolutional layer, six residual blocks, one GRU layer, one fully-connected layer, and an output layer (see Table 1). 
%The stride size of three in the strided convolutional layer is identical to the filter length. 
Residual block comprises two convolutional layers, two batch normalization (BN) layers \cite{BatchNormalization}, two leaky ReLU layers, and a max pooling layer as shown in Table 1. 
Residual connection adds the input of each residual block to the output of the second BN layer. 
A GRU layer with 1024 nodes aggregates frame-level embeddings into an utterance-level embedding. 
One fully-connected layer is used to extract speaker embeddings. 
The output layer has 1211 nodes, which represents the number of speakers in the training set. 
Back-end classifiers including b-vector, rb-vector, concat\&mul comprise four fully-connected layers with 1024 nodes. 

L2 regularization (weight decay) with a weight factor of $10^{-4}$ was applied to all layers. 
An AMSGrad optimizer with learning rate $10^{-3}$ with decay parameter $10^{-4}$ was used \cite{reddi2018convergence}. 
For center loss, $\lambda=10^{-3}$ was used. 
Training was conducted using a mini-batch size of 102. 
Recurrent dropout at a rate of 0.3 was applied to the GRU layer \cite{gal2016theoretically}.

\section{Results and analysis}
Table 2 demonstrates the effectiveness of modifications made to the model architecture (Section 2.1.) and the pre-training scheme (Section 2.2.). 
It also shows the effect of additional objective functions for incorporating explicit feature enhancement phases (Section 2.3.). 
First, modifications to the model architecture and the pre-training scheme reduced the EER from 8.7 \% to 6.8 \%. 
Application of additional objective functions further decreased EER to 4.8 \%. 
The proposed RawNet demonstrates an RER of 44.8 \% compared to the baseline \cite{jung2018avoiding22}. 
\textit{Intermediate}, which does not include additional objective functions, could benefit from explicit feature enhancement techniques. 
On the other hand, RawNet, which includes additional objective functions, exhibits an EER of 4.8 \% without explicit feature enhancement techniques. 
RawNet also outperforms \textit{Intermediate} with feature enhancement showing that it has successfully incorporated the feature enhancement phase. 

Comparison with state-of-the-art front-end embedding extraction systems with cosine similarity back-end is shown in Table 3. 
The proposed RawNet demonstrates the lowest EER compared to both i-vector systems and x-vector systems. 
For i-vector systems, we compared two configurations: with and without LDA feature enhancement. 
For x-vector systems, we compared two configurations based on Shon \textit{et al.} were compared, where the data augmentation scheme introduced in \cite{snyder2018x} was conducted using reverberation and various noise. 

Table 4 describes the performance of various back-end classifiers with the proposed RawNet speaker embeddings. 
In our experiments, PLDA did not show improved results compared to the baseline cosine similarity. 
On the other hand, DNN-based back-end classifiers demonstrated significant improvement, with an RER of 16 \%. 
The rb-vector system did not show additional improvements above the b-vector system. 
Among DNN-based back-end classifiers, the proposed approach of using element-wise multiplication of speaker embeddings from enrol and test utterances performed best, with an EER of 4.0 \%. 

Recent studies using the \texttt{VoxCeleb1} dataset are compared in Table 5. 
The first five rows depict systems that utilize data augmentation techniques. 
``Heavy'' refers to the augmentation scheme of \cite{hajibabaei2018unified, okabe2018attentive}, with various noise from the PRISM dataset and reverberations from the REVERB challenge dataset. 
``Light'' refers to the scheme used in \cite{shon2018frame, snyder2018x} that doubles the size of the dataset using noise and reverberations. 
The i-vector system with PLDA back-end in two implementation versions (one from Nargrani \textit{et al.} and the other from our implementation) demonstrates an EER of 8.8 \% and 5.1 \%, respectively. 
The x-vector system with PLDA back-end without data augmentation, as studied by Shon \textit{et al.}, demonstrates an EER of 7.1 \%. 
The proposed RawNet system with concat\&mul demonstrates the best performance among systems without data augmentation, exhibiting an EER of 4.0 \%. 
The x-vector/PLDA system conducted by Okabe \textit{et al.} \cite{okabe2018attentive} is the only system showing lower EER than our proposed system, but the former is subjected to intensive data augmentation, which hinders direct comparison.

\section{Conclusion}
In this paper, we propose an end-to-end speaker verification system using two DNNs, for extracting speaker embedding extraction and back-end classification. 
The proposed system has a simple, yet efficient, process pipeline where speaker embeddings are extracted directly from raw waveforms and verification results are directly shown using two DNNs. 
Various techniques that compose the proposed RawNet have been explored including the pre-training scheme and additional objective functions. 
RawNet with the concat\&mul back-end classifier demonstrates an EER of 4.0 \% on the \texttt{VoxCeleb1} dataset, which is state-of-the-art among systems without data augmentation, including the x-vector system. 

The proposed RawNet with concat\&mul inputs raw waveforms and outputs verification results. 
Such a simplified process pipeline is expected to lower barriers to research and provide opportunities for many researchers to apply new techniques.

\newpage\newpage
\bibliographystyle{IEEEtran}
\bibliography{mybib}
\end{document}